\documentclass[11pt]{article}
\usepackage{ifthen}
\usepackage{jheppubmod}
\newboolean{elaboratedeclare}
\newboolean{includeapp}
\setboolean{elaboratedeclare}{false}
\setboolean{includeapp}{true}
\pdfoutput=1
\usepackage[showonlyrefs]{mathtools}
\usepackage{psfrag}
\usepackage{array}
\usepackage{amssymb, amsthm}
\usepackage{graphicx}

\usepackage[labelsep=quad]{subcaption}
\usepackage{epstopdf}
\usepackage{url}
\usepackage[hyphenbreaks]{breakurl}

\usepackage{epsfig}

\def\Or[#1]{{\text{O}}\left({#1}\right)}
\def\dotl[#1,#2]{\left\langle #1,\, #2 \right\rangle}
\def\dotlb[#1,#2]{\left\langle #1,\, #2 \right\rangle}
\def\dotlm[#1,#2]{\left[ #1,\, #2 \right]}
\def\dotp[#1,#2]{(\vect{#1} \cdot\vect{#2})}
\def\aff[#1,#2]{\hat{#1}(#2)}
\def\n4sym{{\cal N}=4 SYM}
\def\>{\rangle}
\def\<{\langle}
\def\weight[#1,#2,#3]{\{(#1),#2,#3\}}
\def\ads[#1]{$\text{AdS}_{#1}$}
\def\cft[#1]{$\text{CFT}_{#1}$}

\newcommand{\be}{\begin{equation}}
\newcommand{\ee}{\end{equation}}
\newcommand{\ba}{\begin{align}}
\newcommand{\ea}{\end{align}}
\newcommand{\bs}{\begin{split}}
\def\sess\end{split}

\newcommand{\vect}[1]{{\boldsymbol{#1}}}

\def\pone{\beta_1}
\def\ptwo{\beta_2}
\def\pthree{\beta_3}
\def\ponezer{\beta_1^{(b)}}
\def\ponelock{\beta_1^{(l)}}
\def\ponepost{\beta_1^{(a)}}

\def\kconst{{\cal K}}

\title{Did the Indian lockdown avert deaths?}
\author{Suvrat Raju}
\affiliation{International Centre for Theoretical Sciences, Tata Institute of Fundamental Research, Shivakote, Bengaluru 560089, India.}
\emailAdd{suvrat@icts.res.in}

\abstract{Within the context of SEIR  models, we consider a lockdown that is both imposed and lifted at an early stage of an epidemic. We show that, in these models, although such a lockdown may delay  deaths, it eventually does not avert a significant number of fatalities. Therefore, in these models, the efficacy of a lockdown cannot be gauged by  simply comparing figures for the deaths at the end of the lockdown with the projected figure for deaths by the same date without the lockdown.   We provide a simple but robust heuristic argument to explain why this conclusion should generalize to more elaborate compartmental models. We qualitatively discuss some important effects of a lockdown, which go beyond the scope of simple models, but could cause it to increase or decrease an epidemic's final toll.   Given the significance of these effects in India, and the limitations of currently available data,  we conclude that simple epidemiological models cannot be used to reliably quantify the impact of the Indian lockdown on fatalities caused by the COVID-19 pandemic. 
}

\begin{document}
\maketitle
\section{Introduction}
India entered a nationwide lockdown on 24 March 2020, in response to the Coronavirus disease (COVID-19) pandemic \cite{Chandrashekhar_2020a}. After an initial lockdown period of 21 days, restrictions were gradually eased \cite{Chandrashekhar_2020b}. The lockdown failed to prevent a rapid growth in the number of cases and deaths \cite{kazminft2020}.

On 22 May, the Indian government presented the results of numerous studies that  claimed that the lockdown had averted a significant number of deaths.  One study by the Boston Consulting Group (BCG), which reportedly used a Susceptible-Exposed-Infectious-Recovered (SEIR) model, claimed that the lockdown had saved between 1,20,000 -- 2,10,000 lives \cite{lockdown2020}. Subsequently,  the modelling subgroup of the Indian  Scientists Response to COVID-19 also  announced that it had found, using its model, called ``INDSCI-SIM'', that the lockdown had averted between 8,000 -- 32,000 deaths by 15 May \cite{menon2020}. 

These claims  deserve scrutiny due to the wide publicity that they received and also because they were later invoked
in parliamentary documents and public statements by senior members of the Indian government \cite{nairhindu2020, modijuneaddress2020}.

The purpose of this paper is to explain why, even within the context of simple SEIR epidemiological models, and their generalizations, such claims are misleading. This is because such models uniformly predict that lockdowns that are implemented early in the epidemic---and fail to quash the epidemic---have a {\em negligible} impact on the final death toll. While a lockdown may delay deaths, the models themselves suggests that the fatalities in a ``lockdown scenario'' will rapidly catch up with fatalities in a ``no lockdown scenario.''

Therefore, within the framework of these models, the comparison of fatalities between two scenarios by an arbitrarily chosen fixed date --- in this case, 15 May --- is an absurd metric to gauge the efficacy of a lockdown. This conclusion is described in section \ref{sectheory}.

The conclusions of epidemiological models are known to be often sensitive to the choice of parameters \cite{wu2013sensitivity}. However, the specific property of the models described above is {\em very robust} and insensitive to the parameters chosen. Moreover, while we use a SEIR model to obtain specific analytic results, we explain in section \ref{secheuristic} that our results should qualitatively generalize to more complicated models. Of course, one may choose to eschew models altogether \cite{bhatia2020lessons}. But to the extent that we rely on epidemiological models at all, one of the few reliable lessons we should take away is that lockdowns should not be evaluated by comparing fatalities by a fixed date.  

In the real world, lockdowns may have a positive or negative impact on the epidemic that may be difficult to capture in the model. For instance, if the lockdown quashes the epidemic, or can be used to greatly ramp up healthcare capacity, it may reduce the final toll of the epidemic significantly. On the other hand, if the lockdown engenders economic insecurity, and makes it harder for people to implement long-term precautions like physical distancing, then it may have a negative impact on the eventual toll of the epidemic. We list these effects {\em qualitatively} in section \ref{secrealworld}. Nevertheless, we explain how these effects are much {\em more important} for the final toll of the epidemic than the direct effect of the lockdown itself.  
\ifthenelse{\boolean{includeapp}}{In the Appendix we discuss the INDSCI-SIM model, as an example of an extended SEIR model that confirms our general expectations.}{}

The conclusion of this paper is that simple epidemiological models cannot be used to reliably determine the impact  of the Indian lockdown on fatalities due to the COVID-19 pandemic. 
Rather, an analysis of whether the lockdown has been successful or not must rely on balancing the qualitative factors described in \ref{secrealworld}. It seems apparent to us that a consideration of these factors suggests that the Indian lockdown has contributed to {\em worsening} the final toll of the epidemic.

We would like to state an important caveat at the outset. At various points in this analysis, we use assumptions that are appropriate for the lockdown in India but may not apply to other countries. So the analysis here should not be construed as a comment on lockdowns in other parts of the world.

\section{Lockdowns in the SEIR model \label{sectheory}}

We start by considering the simplest epidemiological models --- the SEIR models. It is possible to obtain simple analytic results in these models. A SEIR model was reportedly used in the BCG study \cite{covidmediamay22}. Moreover, the INDSCI-SIM model can also be thought of as an extended SEIR model.  

However, we will explain in section \ref{secheuristic} why the essential elements of our discussion are robust and model independent, and why we expect them to generalize to other models as well.

\subsection{Setting}
The SEIR model starts by dividing the population into four different compartments. The population is divided into a susceptible group (S), an exposed group (E), an infected group (I) and a removed group (R), which includes those who have recovered as well as those deceased. The differential equations that govern the model are very simple \cite{allen2008mathematical}.
\be
\label{seirequations}
\begin{split}
&{d S \over d t} = {-\pone S I \over N}; \qquad {d E \over d t} = {\pone S I \over N} - \ptwo E;\\
&{d I \over d t} = \ptwo E - \pthree I; \qquad {d R \over d t} = \pthree I.
\end{split}
\ee
We ignore natural births and deaths since they are not relevant for this analysis.

All the variables above are functions of time. But, as above, instead of writing $S(t)$, we will generally suppress this dependence by writing $S$. We only display the time-dependence if we wish to refer to a specific time.  For instance, $S(0)$ is the number of susceptible individuals at time $0$, taken to be the start of the epidemic, and similar conventions hold for the other variables. 

Here, $N$ is the total population size at the start of the epidemic. We assume that $S(0) + E(0) + I(0) + R(0) = N$. This is clearly conserved by the equations. 

In analyzing the model, it is convenient to define a variable $F = I + E$, which satisfies
\be
{d F \over d t} = I \left(\pone {S  \over N} - \pthree \right).
\ee
Moreover, $S$ is itself a monotonically decreasing function in this model, which is clear from the first equation above. It is useful, especially for the analysis of asymptotic properties,
to reparameterize the time-variable in terms of the instantaneous value of the susceptible population. This leads to
\be
{d F \over d S} = \left(\pthree N \over \pone S \right) - 1.
\ee
This can be easily integrated to yield the path of the epidemic in $S-F$ space, which is 
\be
\label{epitrajectory}
{S \over N} - {\beta_3  \over \pone} \log\big({S \over N}\big) + {F \over N} = \kconst.
\ee
Here we have introduced a new constant $\kconst$, which identifies the trajectory of the epidemic.

We may set $\kconst$ using the initial conditions. For instance, if we assume that the epidemic is seeded by a single exposed case, $F(0) = 1$ and that the entire population, except for this individual, is initially susceptible: $S(0) = N-1$. Then we find that $\kconst =  1 -{\beta_3 \over \pone} \log(1 - {1 \over N})$  for the initial conditions above. Since, for a large population, ${1 \over N} \ll 1$, we may simply approximate $\kconst \approx 1$ for this set of initial conditions. But equation \eqref{epitrajectory} is valid for other initial conditions as well.

The constant $\kconst$ is important because it allows us to compute the size of the epidemic as follows. At late times, the epidemic ends and so $F(\infty) = 0$. Therefore the final susceptible fraction, $s_{\infty}=S(\infty)/N$ is just given by solving
\be
\label{epidemicsize}
s_{\infty} - {\beta_3 \over \pone} \log s_{\infty}  = \kconst.
\ee
We will be interested in the case where $\pone > \beta_3$, which is when the epidemic grows exponentially at the start. Then \eqref{epidemicsize} has a single root in the range $s_{\infty} \in (0, {S(0) \over N})$, which is the one we are interested in.

To compute fatalities, we assume a fatality rate, $\mu$. This means that a fraction $\mu$ of all those infected do not recover but instead succumb to the infection. Note that this number is still counted in the variable $R$ and the fatalities, at any point of time, are just $\mu R$. Then the final number of fatalities, which we denote by $D_{\infty}$,  is given by
\[
D_{\infty} = \mu N (1-s_{\infty}),
\]
where $s_{\infty}$ solves the equation \eqref{epidemicsize} above.

We will compare multiple scenarios below. The parameter that we vary in these scenarios is $\pone$ and the quantity we examine is $s_{\infty}$. To avoid confusion, we adopt the convention that whenever these variables appear without a subscript, eg. $\pone$ or $s_{\infty}$, they refer to the generic variable and not a particular value. On the other hand, expressions like $\ponepost$ or $s_{\infty}^{(l)}$ refer to the  values of these variables in a specific scenario.

\subsection{Definition of scenarios}
In this paper, we are concerned with the effect of a lockdown both imposed  and lifted {\em early} in the epidemic.  We make these terms precise below and compare this scenario with a no-lockdown scenario.

\paragraph{\bf Lockdown scenario \\}
In the lockdown scenario, we consider an epidemic that starts with some initial value of $F(0)$ and $S(0) = N-F(0)$. The epidemic starts to  evolve freely  with an initial value of $\pone = \ponezer$. At ${1 - {S \over N}}  = \epsilon_1 \ll 1$, the value of $\pone$ is abruptly changed to $\ponelock$ where $\ponelock < \ponezer$. Then the epidemic evolves with these new parameters until ${1 - {S \over N}} = \epsilon_2 \ll 1$. Note that we always have $\epsilon_2 > \epsilon_1$. Then $\pone$ is changed again to $\ponepost$ such that $\ponelock  < \ponepost < \ponezer$, and then the  epidemic evolves freely till its end.

The interpretation of the scenario above should be obvious. At the beginning, the epidemic evolves with some initial parameters. A lockdown enforces physical distancing and brings down the contact rate between individuals. It does not change the biological parameters of the infection, including its incubation period or time to recovery, and therefore it is modelled by a decrease in $\pone$. When the lockdown ends, the contact rate does {\em not} necessarily return to its initial value. Since several measures, including some measure of physical distancing and perhaps behavioural alterations and other precautions (such as mask-wearing) remain in place after the lockdown,  the model allows for a third value of $\pone$ in the post-lockdown phase. This third value is expected to be in-between the pre-lockdown value, and the value during the lockdown.

The fact that the lockdown is imposed {\em early} and ends early during the epidemic is indicated by $\epsilon_1 \ll 1$ and $\epsilon_2 \ll 1$. 

\paragraph{\bf No-lockdown scenario \\}
In the no-lockdown scenario, we consider an epidemic that starts with the same value of $F(0)$ as above and again take $S(0) = N - F(0)$. The epidemic again initially evolves with  $\pone=\ponezer$. For $1-{S \over N} \in (\epsilon_1, \epsilon_2)$, we take $\beta_1$ to vary  as some monotonic function $\beta_1(1 - {S \over N})$ with the condition that $\beta_1(\epsilon_1) = \ponezer$ and $\beta_1(\epsilon_2) = \ponepost$. The details of this monotonic function will be unimportant for our purpose.  Beyond $1 - {S \over N} = \epsilon_2$, the epidemic evolves with $\pone = \ponepost$ until its end.

The interpretation of this scenario should also be obvious. At the beginning of the epidemic, it evolves with some initial parameters. At some point, behavioural changes are introduced through physical distancing and other precautions. These are the same changes and precautions that set $\pone = \ponepost$ in the lockdown scenario. These changes can often be implemented rapidly and, in fact,  they can be implemented instead of the lockdown.  So it would be quite reasonable to approximate the function $\beta_1(1-{S \over N})$ by a step function with a term proportional to $\theta(1 - {S \over N} - \epsilon_1)$. Nevertheless, we consider a more general scenario where these behavioural changes and precautions are implemented more gradually. But we do assume that they are in place by the time the epidemic has reached the  stage, $1-{S \over N} = \epsilon_2$, which is the stage of susceptibility where the lockdown {\em ends} in the ``lockdown scenario''.

However, note, in particular, that the rate, $\pone$, in the lockdown scenario is {\em always lower or equal}   to  its value in the no-lockdown scenario. 

\subsection{Effect of a lockdown on fatalities}
We now compare final fatalities in the lockdown and no-lockdown scenarios. 

\paragraph{Fatalities in the lockdown scenario \\}
The change in parameters, upon the imposition of the lockdown, also leads to a 
change in the value of $\kconst$ by $\delta \kconst_{11}$, which is given by 
\be
\delta \kconst_{11} = \left({1 \over \ponezer}  - {1 \over \ponelock} \right) \pthree \log(1-\epsilon_1).
\ee
This equation is derived by noting that $S$ and $F$ remain continuous when the lockdown is imposed, and the change in $\kconst$ comes only from the change in the value of $\pone$. 
When the lockdown is lifted, there is an additional change in this constant, $\delta \kconst_{1 2}$, which is given by
\be
\delta \kconst_{12} = \left({1 \over \ponelock}  - {1 \over \ponepost} \right) \pthree \log(1-\epsilon_2).
\ee

The net change in $\kconst$ can be expanded out to first order in $\epsilon_1$ and $\epsilon_2$ as
\be
\delta \kconst_{1} = \delta \kconst_{11} + \delta \kconst_{12} = -\pthree \left(\frac{\epsilon_2-\epsilon_1}{\ponelock}+\frac{\epsilon_1}{\ponezer}-\frac{\epsilon_2}{\ponepost}\right)+ {\cal S}_1,
\ee
where we have clubbed the higher order terms into ${\cal S}_1$, since they will be largely  irrelevant for our analysis.
\be
{\cal S}_1 = -\frac{\pthree  \left(\epsilon_1^2 \ponepost (\ponelock-\ponezer)+\ponezer \epsilon_2^2
   (\ponepost-\ponelock)\right)}{2 \ponelock \ponezer \ponepost}+O\left(\epsilon^3 \right).
\ee

It is convenient to compare the final toll of the epidemic by comparing it with a ``reference'' epidemic, where the value of $\pone = \ponepost$ throughout the epidemic, and where the epidemic starts with the same initial value of $\kconst$ as the epidemic above.  If we define the solution of \eqref{epidemicsize} for this reference case  as $s_{\infty}^{\text{ref}}$ and if we denote the value of $s_{\infty}$ in the lockdown scenario by $s_{\infty}^{(l)}$, then we can then use equation \eqref{epidemicsize} to compute the difference between these two quantities.

To first order, this is controlled by
\be
{d s_{\infty} \over d \kconst} = -\left({\pthree \over \ponepost s_{\infty}} - 1\right)^{-1},
\ee
where the quantity in brackets is positive. And therefore
\be
 s_{\infty}^{(l)} - s_{\infty}^{\text{ref}} = \left({\pthree \over \ponepost s_{\infty}^{\text{ref}}}  - 1\right)^{-1} \pthree \left(\frac{\epsilon_2-\epsilon_1}{\ponelock}+\frac{\epsilon_1}{\ponezer}-\frac{\epsilon_2}{\ponepost}\right) + \Or[\epsilon^2].
\ee

\paragraph{Fatalities in the no-lockdown scenario \\}

In the ``no lockdown'' scenario, we have an unknown function $\pone(1 - {S \over N})$ which controls how fast the long-term equilibrium value of $\pone = \ponepost$ is achieved. Even though this function is unknown, under the assumption that it is monotonic, we can {\em bound} the asymptotic value of $s_{\infty}$ both above and below by considering two extreme subcases within this scenario. 

First consider the case where the value of $\pone$ changes abruptly to $\ponepost$ when ${1 - {S \over N}} = \epsilon_1$. In this case, the epidemic shifts to a new trajectory where $\kconst$ changes by
\be
\delta \kconst_{21} = \beta_3 \log(1 - \epsilon_1) \left( {1 \over \ponezer}  - {1 \over \ponepost}\right) = -\beta_3 \epsilon_1 \left( {1 \over \ponezer}  - {1 \over \ponepost}\right) + \Or[\epsilon^2].
\ee
The second sub-scenario is the case where the value of $\beta_1$ stays constant until ${1 - {S \over N}} = \epsilon_2$ and then changes abruptly to $\ponepost$. In this case, the value of $\kconst$ changes by 
\be
\delta \kconst_{22} = \beta_3 \log(1 - \epsilon_2) \left({1 \over \ponezer} - {1 \over \ponepost} \right) = -\beta_3 \epsilon_2 \left( {1 \over \ponezer}  - {1 \over \ponepost}\right) + \Or[\epsilon^2].
\ee

The reason to consider these two sub-scenarios is that the final value of $s_{\infty}$ in the ``no lockdown'' scenario is bounded on both sides by its value in the cases above. In particular, denote the final value of the susceptibility in an arbitrary no-lockdown scenario (as specified above) by $s_{\infty}^{(n)}$. Then, we find that independent of the details of the variation of $\beta_1(1 - {S \over N})$ in the range ${1 - {S \over N}} \in (\epsilon_1, \epsilon_2)$ in this scenario, we have
\be
s_{\infty}^{\text{ref}} - s_{\infty}^{(n)} \geq s_{\infty}^{\text{ref}} - s_{\infty}^{(n_1)} = \left({\pthree \over \ponepost s_{\infty}^{\text{ref}}}  - 1\right)^{-1} \pthree \epsilon_1 \left({1 \over \ponepost} - {1 \over \ponezer}\right).
\ee
We also have
\be
s_{\infty}^{\text{ref}} - s_{\infty}^{(n)} \leq s_{\infty}^{\text{ref}} - s_{\infty}^{(n_2)} = \left({\pthree \over \ponepost s_{\infty}^{\text{ref}}}  - 1\right)^{-1} \pthree \epsilon_2 \left({1 \over \ponepost} - {1 \over \ponezer}\right).
\ee
Here $s_{\infty}^{(n_1)}$ and $s_{\infty}^{(n_2)}$ are the final susceptible fractions in the two extreme no-lockdown scenarios.

\paragraph{\bf Difference in fatalities \\}
The difference in fatalities in the two scenarios is proportional to this difference in $s_{\infty}$ in the two scenarios. We have
\be
\delta D = \mu N (s_{\infty}^{(l)} - s_{\infty}^{(n)}).
\ee
Even though we have an unknown interpolating function in the no-lockdown scenario we can {\em bound} $\delta D$ both from above and below. We find that the factors conspire to give the simple result
\be
\label{deathsaverted}
{\mu N \pthree \over \left({\pthree \over \ponepost s_{\infty}^{\text{ref}}}  - 1\right)}  (\epsilon_2 -  \epsilon_1) \left({1 \over \ponelock} - {1 \over \ponepost}\right)
\leq \delta D \leq {\mu N \pthree \over \left({\pthree \over \ponepost s_{\infty}^{\text{ref}}}  - 1\right)}  (\epsilon_2 -  \epsilon_1) \left({1 \over \ponelock} - {1 \over \ponezer}\right).
\ee
We will use the notation $(\delta D)^{(n_1)}$ for  the lower bound above and $(\delta D)^{(n_2)}$ for the upper bound. These correspond to the difference in deaths between the lockdown scenario and the two extreme no-lockdown scenarios detailed above.

The reason that this effect is small is because it is proportional to $\epsilon_1$ and $\epsilon_2$ and, by assumption, these numbers are small for the ``early lockdown'' scenario that we consider. 

 \paragraph{\bf Deaths averted and deaths during lockdown \\}
 In fact, a useful estimate of the figure of deaths averted is given by the number of deaths that take place {\em during} the lockdown.

 Note that the fatalities observed during the lockdown can be computed from \eqref{epitrajectory} to be
 \be
\delta D^{(l)} = \mu N {\pthree \over \ponelock} \delta (\log(S)) \approx \mu N {\pthree \over \ponelock} (\epsilon_2 - \epsilon_1) + \Or[\epsilon^2]. 
 \ee
In particular we see that
\be
\label{boundingratio}
  {1 - {\ponelock \over \ponepost} \over \left({\pthree \over \ponepost s_{\infty}^{\text{ref}}}  - 1\right)}
 \leq {\delta D \over \delta D^{(l)}} \leq {1 - {\ponelock \over \ponezer} \over \left({\pthree \over \ponepost s_{\infty}^{\text{ref}}}  - 1\right)}.
\ee
Note that several uncertain parameters --- including the fatality rate, $\mu$, and even $\epsilon_1$ and $\epsilon_2$ --- have dropped out of the formula above.

The $\Or[1]$ factors that remain above depend, in detail, on the various constants in the problem. In fact, for several reasonable values of the parameters, these factors are {\em smaller than 1} meaning that deaths averted are smaller than the deaths that have taken place during the lockdown. But, in any case, an immediate implication is that one should be skeptical of estimates that suggest that deaths averted are much higher than deaths during the lockdown. In India's case, since 2743 deaths had been recorded during the lockdown, from 24 March to 15 May \cite{covidjhudb}, this immediately tells us that the significantly higher figures described in the Introduction are absurd, even within the context of epidemiological models.

Here, we would like emphasize an important general point. In the SEIR model, the figure for deaths averted by the lockdown is  {\em entirely separate} from the figure for the deaths that take place in the same {\em time period} in the no-lockdown scenarios. If one considers an ``effective lockdown'' that is imposed early, then even after a long time (in units set by $\pone^{-1}, \ptwo^{-1}, \pthree^{-1}$) the value of $\epsilon_2$ may remain very small. In the meantime, the deaths in the no-lockdown scenario in the same time period will diverge exponentially from the deaths in the lockdown scenario. It is this exponentially large difference that gives rise to the figures described in the Introduction. But, within the model, this  large number is irrelevant for the final figure of the deaths averted.

\subsection{A numerical example}
The results above might seem puzzling, since ``common sense'' suggests that a lockdown should ``avert deaths''. Before we explain them heuristically, we provide a simple numerical example.

The point of this example is {\em not} to make realistic predictions for the Indian epidemic. The model is too simple to do that, and moreover the available data---including that of the total number of cases and fatalities---is too flawed. Our point is only to illustrate the theoretical effect described above. For this reason, we deliberately do {\em not consider} the fatality data for India. 

Instead we consider a hypothetical epidemic in a population of $N=10^7$. The epidemic is seeded by a small set of exposed cases: $E(0) = 10, I(0) = 0, S(0) = N-10$.  The fatality rate is taken to be $\mu = .005$ or $0.5\%$. We assume that $\beta_2 = {1 \over 3} (\text{day})^{-1}$ and $\beta_3 = {1 \over 5} (\text{day})^{-1}$. In our hypothetical example,  before the lockdown, the epidemic grows rapidly with $\ponezer = {3 \over 5} (\text{day})^{-1}$. In the lockdown scenario, a lockdown of 70 days is imposed on day 20 of the epidemic and lifted on day 90 of the epidemic. The lockdown brings $\pone$ down to $\ponelock = {3 \over 10} (\text{day})^{-1}$.  This lockdown corresponds to the values of $\epsilon_1 = 4.8 \times 10^{-5}$ and $\epsilon_2 = 4.9 \times 10^{-3}$ and therefore our assumption of an ``early lockdown'' is met. At the end of the lockdown, $\pone$ rises again to $\ponepost={2 \over 5} (\text{day})^{-1}$. It does not rise to its original value, since some sustainable behavioural changes are made and some long-term precautions are put in place.

We compare this lockdown scenario with the two ``no lockdown'' scenarios described above. In one scenario, denoted by ``NL1'' in the graphs below, the long-term precautions  changes are implemented, {\em instead} of the lockdown on day 20. In the other scenario, denoted by ``NL2'', the epidemic continues to evolve freely past day 20. The long-term precautions are implemented only when  ${1 - {S \over N}} = \epsilon_2$. In terms of time, this value of $\epsilon_2$ is reached at $44.90$ days. Therefore, in the scenario NL2, the  long-term precautions and behavioural changes that bring $\pone$ down from $\ponezer$ to $\ponepost$ are put in place 25 days after the lockdown is imposed in scenario ``L''.   As explained above, if we take a more general variation of $\pone$ we expect the asymptotic fatalities to be bounded above and below by these two scenarios.

\begin{figure}[!h]
\begin{center}
\includegraphics[width=0.6\textwidth]{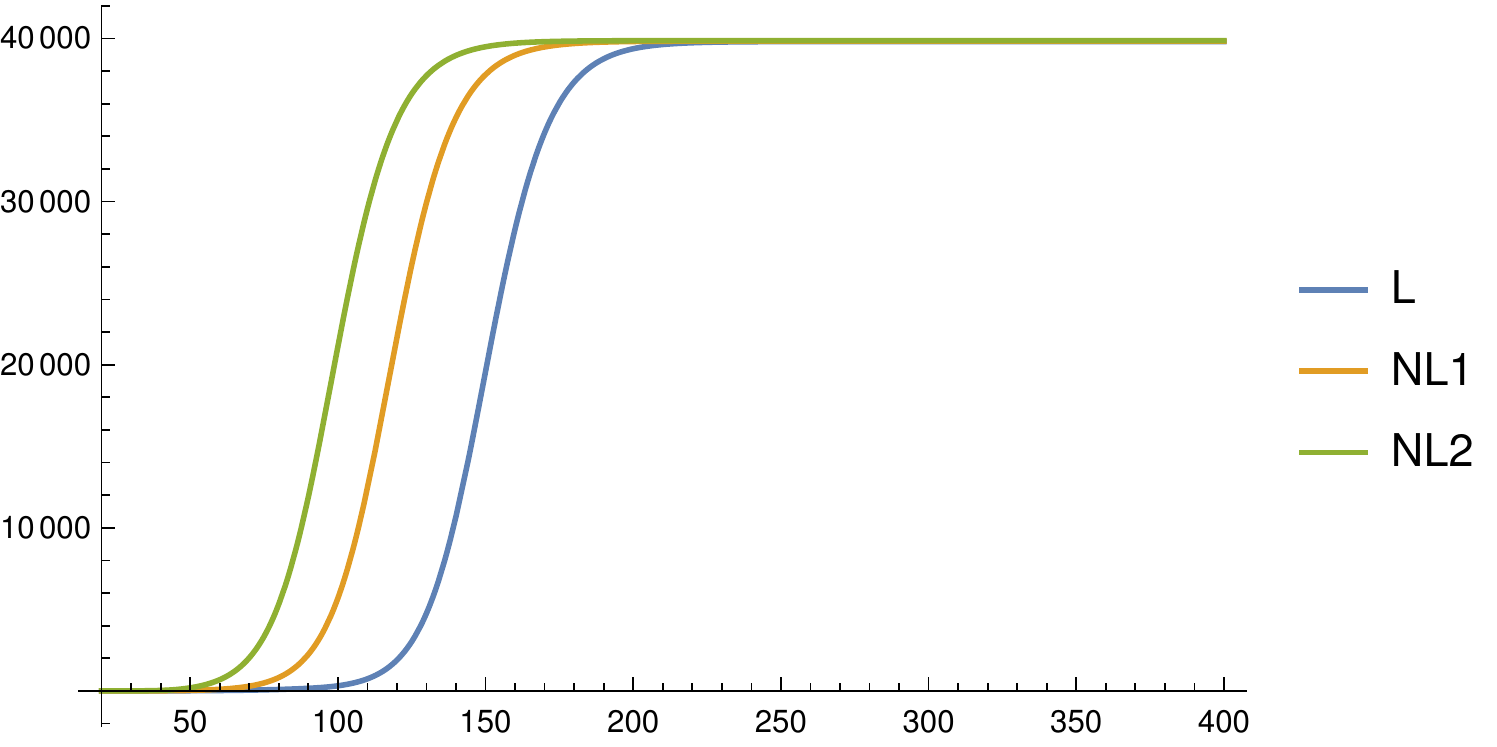}
\caption{\em Total deaths as a function of time in the three scenarios. ``L'' is the lockdown followed by long-term precautions; NL1 implements only the long-term precautions instead of the lockdown; NL2 implements long-term precautions after a delay.  \label{figfatalities}}
\end{center}
\end{figure}

The fatalities as a function of time in the three scenarios are shown in Figure \ref{figfatalities}.
This figure perfectly bears out the analysis above. When one compares the lockdown scenario to NL2 scenario on day 90, the no-lockdown scenario has had 11,722 excess deaths.  However, it is absurd to conclude from this figure that the lockdown has ``averted more than 11,000 deaths''. This is because, at the end of 365 days, deaths in the lockdown scenario have caught up. In fact, in the model, there are less than 56 excess deaths after 365 days compared to the lockdown scenario.

The comparison with the NL1 scenario is similar. Here, in the model, there are 2078 excess deaths compared to the lockdown scenario by day 90. But after 365 days, the difference in deaths drops below 28.

\begin{figure}[!h]
\begin{center}
\includegraphics[width=0.6\textwidth]{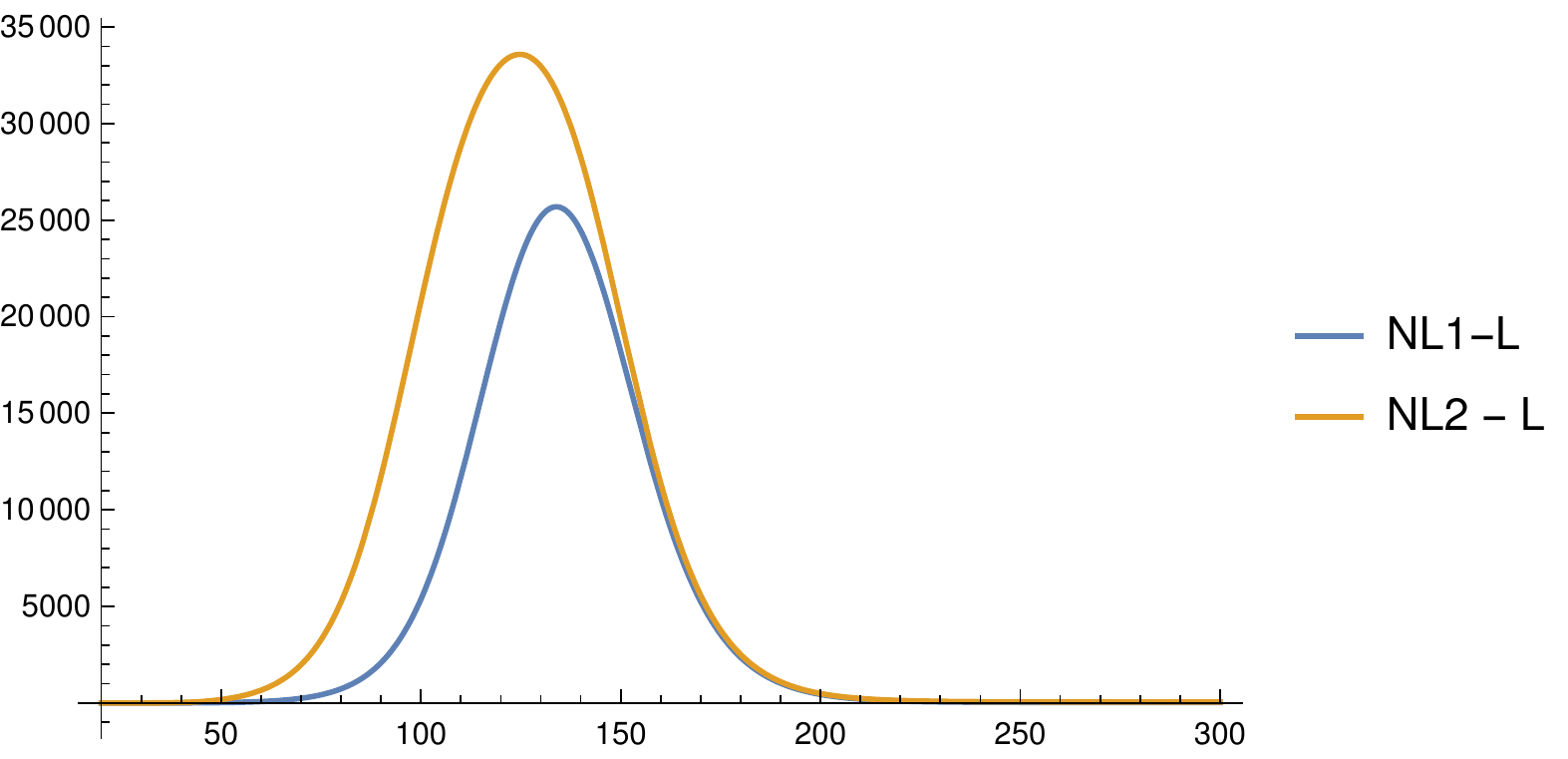}
\caption{\em Death delayed by the lockdown. The graph shows the difference in fatalities between the lockdown scenario and the no-lockdown scenarios as a function of time.
 By the end of the epidemic, deaths in the lockdown scenario catch up almost entirely with the no-lockdown scenarios. \label{figdelayed}}
\end{center}
\end{figure}

The difference in deaths between the lockdown scenario and the NL1 and NL2 scenarios is also plotted in Figure \ref{figdelayed}.
This graph shows that the difference in deaths appears to become very large at intermediate times, but then drops to a very small number  by the end of the epidemic.

The example also provides us with an opportunity to check our first order analytical formulas. The match between the numerics and the first order approximations is very good as shown in the table below.
\[
\begin{array}{|c|c|c|c|}
\hline
\text{Quantity}&\text{Numerical} (N) &\text{First-order Approx} (A) & \text{Error} \left({|N-A| \over N}\right) \\
\hline
\delta \kconst_{1} &-8.06 \times 10^{-4} & -8.04 \times 10^{-4} &2.5 \times 10^{-3} \\
\delta \kconst_{21} &8.03 \times 10^{-6} &  8.03 \times 10^{-6} &2.4 \times 10^{-5} \\
\delta \kconst_{22} &8.22 \times 10^{-4} & 8.20 \times 10^{-4} &2.5 \times 10^{-3} \\
\delta D^{(l)}  &27.64 & 27.51  & 4.8 \times 10^{-3}
\\
\delta D^{(n_1)}  &27.92 & 27.78 & 4.7 \times 10^{-3} \\
\delta D^{(n_2)}  &55.70 & 55.57 & 2.4 \times 10^{-3} \\
\hline
\end{array}
\]
We also note that for these parameters, the lower and upper bounds that appear in \eqref{boundingratio} are 0.17 and 0.34 respectively. In this model, the deaths that take place during the lockdown are given by $\delta D^{(l)} = 162.75$. As expected,  the two extreme no-lockdown solutions saturate the lower and upper bound of \eqref{boundingratio}. So, in this example, the deaths averted are significantly {\em smaller} than the deaths that take place during the lockdown.

The reader might wonder, from Figure \ref{figfatalities} whether, in the lockdown scenario, the peak of infections is ``flattened''. However, even this does not happen. For the parameters above, although the peaks occur at significantly different times, the actual peak {\em values} of $I$ in the NL2, NL1 and L scenarios are similar and given by
\be
I_{\text{max}}^{(n_2)} = 9.56 \times 10^5; \quad  I_{\text{max}}^{(n_1)} = 9.51 \times 10^{5}; \quad I_{\text{max}}^{(l)} =  9.46 \times 10^5.
\ee
So, we have
\be
{I_{\text{max}}^{(n_2)} - I_{\text{max}}^{(l)} \over N}  = 1.0 \times 10^{-3}; \qquad {I_{\text{max}}^{(n_1)} - I_{\text{max}}^{(l)}  \over N} = 5.0 \times 10^{-4}.
\ee
This is entirely consistent with the understanding that we develop in section \ref{secheuristic}. Although the focus of this note is on ``deaths averted'' in fact, in reasonable models, all the {\em intrinsic properties} of the trajectory with an early lockdown are very close to those of the trajectory without the lockdown.

\section{A heuristic explanation and general compartmental models \label{secheuristic}}

Epidemiological models are often unreliable and subject to sensitivity in the parameters. In fact, this author is of the opinion that the {\em only} predictions of models that should be regarded as robust and taken seriously are those that can {\em also be reproduced} by a simple heuristic argument. So in this section we provide a heuristic argument to explain the result above, which states that a lockdown has only a small impact on the final number of deaths. We then use our heuristic explanation to generalize our result to extensions of the SEIR model.

\subsection{The SEIR Model}
The formula \eqref{deathsaverted} has two aspects. The first aspect, which is the important one qualitatively, is that the result for deaths averted is $\Or[\epsilon]$. This means that, in the limit, where we take the lockdown to be imposed and lifted earlier and earlier in the epidemic, its effect on the final number of fatalities vanishes. This is easy to understand. In such a situation, the lockdown only {\em translates} the entire curve of infections to the right, without having any appreciable impact on its shape. In particular, at leading order, an early lockdown does {\em not avert fatalities} at all.

This is a robust argument, and independent of the details of the model. Therefore we expect that, even in more complicated models, lockdowns, by themselves, should be ineffective in reducing deaths. 

The formula \eqref{deathsaverted} also has an $\Or[\epsilon]$ piece and this term can also be understood through a slightly more elaborate heuristic argument. The key point is to track the epidemic's progress in terms of the fraction of the population that remains susceptible, rather than directly use the time-variable. Even in more complicated models, with a number of compartments, as long as recovery from infection grants immunity, the fraction of the population that remains susceptible is a monotonic function and so it can be used to track the progress of the epidemic.

Let us use this to understand how the results obtained above for the SEIR model can be understood heuristically. In this model, at the start when $1 - {S \over N} \approx 0$, each individual effectively infects ${\ponezer \over \pthree}$ other individuals. In one of the extreme no-lockdown scenarios above, when  $1 - {S \over N} = \epsilon_2$, this number is lowered to ${\ponepost \over \pthree}$.   The epidemic grows until ${S \over N} = {\pthree \over \ponepost}$. Beyond this threshold, the epidemic starts to decay but it does not end.  Importantly the final toll of the epidemic is controlled by the threshold where it starts to decay but {\em also} the number of exposed and infected individuals at this threshold. These individuals cause an ``overshoot'' beyond the ``herd immunity'' threshold and increase the final toll of the epidemic.

Now the lockdown alters this trajectory as follows. First note that since it takes longer for the population to reach the level $1 - {S \over N} = \epsilon_2$, the lockdown tends to advance recoveries over the no-lockdown scenario. This effect is clearly proportional to $\epsilon_2 - \epsilon_1$. 
But this implies that the total number of infected and exposed individuals are smaller in the lockdown scenario than in the no-lockdown scenario, when the two epidemics are compared at the same value of $1 - {S \over N}$. This  effect is proportional to $(\epsilon_2 - \epsilon_1)$ since $R+F+S=N$ and so the change in $F$ is precisely the change in $S$. This effect causes the epidemic to end at a lower value of ${S \over N}$ than without the lockdown. 
Since the final number of fatalities is controlled by the end-point of the epidemic, one sees that the lockdown lowers fatalities by a term that is proportional to $\epsilon_2 - \epsilon_1$. 

The comparison of the lockdown scenario with the other extreme no-lockdown scenario --- where behavioural changes are implemented {\em instead} of the lockdown and at the same time as the lockdown--- is similar. One again finds that fatalities in the lockdown scenario are lower by a factor that is proportional to $\epsilon_2 - \epsilon_1$. 

\subsection{General compartmental models \label{generalcompart}}
We now explain why we expect the heuristic argument above to continue to apply to more general compartmental models.

In more complicated models, the population is often divided into additional compartments. For instance, instead of simply considering ``infected individuals'' one may divide them into a group that is ``asymptomatic'' and another that is ``symptomatic''.  The details of these compartments are not important for our argument and say that we have divided the population into $Q$-compartments, whose occupancy we denote by ${\cal C}_{i}$, with $i = 1 \ldots Q$. 

The dynamics associated with the additional compartments will also be different from the SEIR model. For instance,  ``asymptomatic'' and ``symptomatic'' individuals might pass on the infection to susceptible individuals at different rates. The details of these dynamics are also not important for the argument,  and say that the trajectory of the epidemic without the lockdown follows a path
\be
\label{nolock}
{d {\cal C}^{(n)}_{i}  \over d S} = {\cal F}_i ({S \over N}, {{\cal C}^{(n)}_i \over N}),
\ee
for some set of functions ${\cal F}_i$. These functions are $\Or[1]$ functions of $\Or[1]$ quantities and not expected to scale with $N$ since the left hand side does not scale with $N$ either. This represents the fact that in such models, the important quantity is the {\em fraction} of the total population that is in a compartment and not the absolute number for the occupancy of that compartment.

The framework described by \eqref{nolock} is quite general. For instance,  it can be used to easily model a fatality rate that is a function of the instantaneous number of infections or the occupancy of another compartment in the model. To account for this possible effect, we introduce a compartment for fatalities, which we call ${\cal C}_{D}$. In the SEIR model, with a constant death-rate, this would always just be proportional to recoveries, but if the fatality rate depends on other variables, it might be different.

 Note also that, in writing the equations in the form \eqref{nolock}, where the trajectory is parameterized by the susceptibility, we have assumed that there is no {\em explicit} time-dependence in the parameters that govern the model. This assumption may be false if, for instance, the healthcare system undergoes a rapid improvement over a short period of time.  We return to this issue in section \ref{secrealworld}.

Now, a lockdown alters the dynamics of the epidemic for a limited period. This means that the lockdown changes the trajectory of the epidemic to one that satisfies 
\be
\label{lock}
{d {\cal C}_{i}^{(l)}  \over d S} = {\cal F}_i({S \over N}, {{\cal C}_i^{(l)} \over N}) + \theta(1 - {S \over N} - \epsilon_1) \theta(\epsilon_2 - 1 + {S \over N}) {\cal G}_i ({S \over N}, {{\cal C}^{(l)}_i \over N}).
\ee
Here, during the lockdown, which starts when  the fraction of the susceptible population is $1 - \epsilon_1$ and lasts till it becomes $1 - \epsilon_2$,  we have assumed that the trajectory is controlled by some new evolution functions that differ from the old ones by  ${\cal G}_i$.  

One may choose to set the start and end-points of the lockdown in terms of time, rather than susceptibility, but equation \eqref{lock} still applies with $\epsilon_1$ and $\epsilon_2$ merely being set by the value of $1 - {S \over N}$ at the starting and ending points of the lockdown. 

Since the impact of the lockdown is constrained to  the interval $1 - {S \over N} \in  (\epsilon_1, \epsilon_2)$, the trajectories \eqref{nolock} and \eqref{lock} satisfy the {\em same} differential equation both before and after this interval.  In particular, if we compare the trajectories of the two scenarios after the lockdown, then it is only the initial conditions for these two trajectories that  are different. Moreover, these initial conditions differ by only a small amount. This can be written as
\be
\label{boundeddiff}
{1 \over N} \left|{\cal C}_{i}^{(l)}(S = N (1 - \epsilon_2)) - {\cal C}_i^{(n)}(S=N(1 - \epsilon_2)) \right| =   (\epsilon_2 - \epsilon_1) \rho_i,
\ee
where $\rho_i$ is some set of model-dependent $\Or[1]$ constants controlled by the manner in which the lockdown modifies dynamics.

In the SEIR model the ``termination condition'' for the epidemic was simply $F = 0$. Say that this termination condition generalizes to 
\be
\label{stopping}
{\cal T}(S, {\cal C}_i) = 0,
\ee
in the more elaborate model. 
We are interested in the value of  $S$, at which \eqref{stopping} is satisfied and the expected fatalities can then be computed as ${\cal C}_{D}(S)$. But as a result of \eqref{boundeddiff}, we expect that if \eqref{stopping} is solved by some value $S^{(l)}$ for the lockdown trajectory and some other value $S^{(n)}$  for the no-lockdown trajectory then
\be
 {{\cal C}_{D}(S^{(n)}) \over N}  - {{\cal C}_{D}(S^{(l)}) \over N}  = (\epsilon_2 - \epsilon_1) \sigma,
\ee
where $\sigma$ is again some $\Or[1]$ constant that depends on the model. 

The assumption here is that the differential equations that govern the epidemic,  \eqref{nolock}, are {\em not chaotic} so that a small difference in the initial conditions, \eqref{boundeddiff}, leads to a small difference in the asymptotic value of ${S \over N}$. This is a reasonable assumption to make about epidemiological models.\footnote{As discussed in section \ref{secrealworld} there are some exceptions to this assumption.  For instance, it may happen that when the ${\cal C}_i$ are below some threshold, the healthcare system can keep them below that threshold indefinitely and not otherwise. In such a case, small differences in initial conditions could make a significant difference in the final outcome. These exceptions, which apply when the epidemic is quashed by the lockdown, do not apply to the Indian lockdown as explained in section \ref{secrealworld}.}

To summarize:  while the precise factors in \eqref{deathsaverted} are specific to the simple SEIR model, in more complicated models,  we still expect that the number of deaths averted by a lockdown imposed and lifted early in the epidemic will be $\Or[\epsilon]$ 

While we have paid special attention to the question of fatalities averted, the argument above is general enough that it can be applied to {\em any intrinsic} property of the trajectory. For instance, one may introduce a compartment corresponding to hospitalized patients, and ask about the peak occupancy of this compartment.  But the argument above tells us that, when parameterized by the susceptible population, the entire trajectory of the lockdown scenario is separated from the no-lockdown scenario by a difference proportional to $(\epsilon_2 - \epsilon_1)$.

Therefore, the argument implies that the difference between this peak, in the lockdown and no-lockdown scenarios, will also be proportional to $(\epsilon_2 - \epsilon_1)$. An early lockdown may {\em delay} the peak in terms of time, but it will not bring down its height. \ifthenelse{\boolean{includeapp}}{In Appendix \ref{appisim} we verify some of these predictions for one example of an extended SEIR model.}{} 

In this section, we have heuristically argued and verified that even in general compartmental models, the effect of a lockdown is small. The reader might correctly point out that if the population is large then even an $\Or[\epsilon]$ fraction might translate into a large number, in terms of actual lives. However, it is important to understand that these $\Or[\epsilon]$ predictions made by models are not reliable. As we explain in section \ref{secrealworld}, the real-world impact of a lockdown is controlled by different factors, which are not accounted for in simple models, but can change the outcome of the epidemic much more significantly than these $\Or[\epsilon]$ effects.

\section{Real-world caveats \label{secrealworld}}
The results above apply in an idealized setting, and so we now describe several real-world caveats to the results. In reality, a lockdown may have both negative and positive impacts on the final toll of an epidemic. We list some of the factors that could lead to this. We are especially interested in those factors that might be relevant in India's case.

\subsection{Possible negative long-term impacts of a lockdown}
A lockdown may have a number of negative long-term impacts. 

\begin{enumerate}
\item{\bf Lockdown leading to larger long-term values of $\pone$ \\} 
We described above how the eventual toll of the epidemic is almost entirely controlled by its rate of spread after the lockdown ends; we denoted the relevant parameter by $\ponepost$ above. In reality, $\ponepost$ depends on {\em long-term sustainable precautions} that individuals can take. A lockdown that fails to emphasize welfare --- as has been the case in India --- may make it harder for people to implement such precautions.

For instance, measures such as reducing the number of working days in a week, or implementing staggered hours for markets and shops to reduce crowding can contribute to a reduction of $\ponepost$. However, such precautions almost invariably have an economic cost. If the lockdown has engendered economic insecurity, and chipped away at the economic and social reserves of people, it may make it harder to implement such precautions sustainably in the long term. 

Therefore, the lockdown, by itself, {\em may cause an increase} in the long term value of $\pone$, by making it harder for people to implement sustainable physical-distancing measures. This can easily overwhelm the very small $\Or[\epsilon]$ direct gains in deaths averted due to the lockdown.

\item{\bf Lockdown leading to a humanitarian crisis \\} 
  More than 90\% of India's workforce is employed in the unorganized sector, or informally employed in the organized sector \cite{economicsurvey1819}. Consequently, the economic impact of the lockdown has been severe. The Center for Monitoring Indian Economy estimated that, despite a slightly improvement in the month of May compared to the month of April, ``over a 100 million people were still out of jobs compared to employment in 2019-20'' \cite{cmiejun2020}.  Another glimpse of the extent of the crisis is provided by the observation made by the Stranded Workers Action Network (SWAN). SWAN contacted thousands of workers but found that  ``90\% $\ldots$  did not get paid by their employers'' and  ``96\% $\ldots$ did not get rations from the government'' \cite{swan2020}. 

This humanitarian crisis is a direct consequence of the lockdown, and constitutes the single most significant impact that the lockdown has had on India's people.

\item{\bf Lockdown reducing access to healthcare \\}
A lockdown may make it harder for people to access healthcare for other conditions. It is known that tuberculosis (TB), by itself, claims about 450,000 lives every year in India \cite{globaltbreport19}. But, as a direct consequence of the lockdown, reports suggest that the number of new TB cases notified in government healthcare centers fell very sharply in April 2020 when compared with the previous year \cite{beditb2020}.
 If the Indian lockdown causes the rate of TB deaths to rise by even a fraction, it could again easily overwhelm the $\Or[\epsilon]$ gains made in combating COVID-19.

\end{enumerate}

\subsection{Possible positive long-term impacts of a lockdown}
A lockdown may possibly have positive long-term effects although it is not clear if such effects are directly visible in India's case.
\begin{enumerate}
\item
{\bf Possibility of quashing the epidemic \\}
A lockdown may succeed in quashing the epidemic or reducing the epidemic to a very low level so that, through testing, tracing and other interventions, its rate of spread can subsequently be kept low for a long period. When the number of cases is very low, the dynamics of an epidemic cannot be reliably understood through compartmental models. For instance, at the time of writing, the lockdown in New Zealand is believed to have successfully limited COVID-19 transmission there \cite{Baker_Wilson_Anglemyer_2020}.  Similarly, the lockdown in Hubei that provided the template for other lockdowns, brought the epidemic down to such a low level that  testing-and-tracing has subsequently succeeded in keeping the number of new infections very limited \cite{Leung_Wu_Liu_Leung_2020}. 

This is irrelevant for the discussion in India, where the epidemic has evidently not been quashed by the lockdown.
\item
{\bf Late lockdowns \\}
If a lockdown is imposed late in the epidemic, when the number of new infections is close to its peak, then even a simple lockdown  may reduce the number of fatalities by reducing the ``overshoot'' beyond herd immunity. In this situation, the analysis of section \ref{sectheory} would  not be  valid since $\epsilon_1, \epsilon_2$ would not be small.
\item
{\bf Preparing the healthcare system \\}
A lockdown may reduce the absolute number of cases, and so reduce the immediate stress on the healthcare system. It may also provide time for the state to ramp up health facilities. We use this term to include the ramping-up of testing and quarantine facilities.

We note that the observation that the lockdown has {\em currently} prevented healthcare facilities from being overwhelmed does not, by itself, provide a compelling rationale for the lockdown.  This  involves the same logical error that we described above in the case of fatalities. If the lockdown merely {\em delays} the date on which the healthcare system is overwhelmed, then this implies that its long-term positive-effects are limited.

Therefore, the question that must be asked is whether the lockdown has been used to ramp up health facilities so that they are better-prepared to confront the epidemic in the long run.

\item
{\bf Behavioural changes \\}
It is sometimes suggested that the lockdown was necessary  to promote behavioural change and to persuade individuals to  adopt precautions \cite{lee2020job}.  We list this factor for completeness but we do not believe that this constitutes a genuinely long-term impact. For instance, while wearing masks  in public or hand-washing where possible are both important precautions, it appears clear that the these precautions could have been encouraged even without the lockdown.

\item{\bf A Vaccine \\}
If a vaccine for a disease is already available, the delay induced by a lockdown can, in principle, be used to vaccinate a large number of people. However, in the case of the COVID-19 pandemic, in spite of early encouraging results \cite{bar2020encouraging} it appears clear that a vaccine will not be available at least for several months and perhaps for longer. Moreover, even after a  safe vaccine becomes available, it will take a considerable amount of time to actually vaccinate a significant fraction of the population in India. Since these time-scales are so much longer than the time-scale of the delay induced by the Indian lockdown, the potential availability of a vaccine in the future cannot be used to evaluate the lockdown in India.
\end{enumerate}

The objective of this paper is not to enter into a detailed discussion of these qualitative effects.  Here, we would merely like to point out that the factors above are {\em far more important} than meaningless measures of deaths-delayed by a fixed date. This is because even small changes in the factors above --- either in the positive or the negative direction ---  can completely overwhelm the $\Or[\epsilon]$ fatalities averted by the lockdown. So the lockdown must be gauged by evaluating whether the positive long-term effects above outweigh its negative long-term effects.

In principle, some of the effects above can be included as parameters in models.  For instance 
one could include a ``back-reaction effect'' of the lockdown accounting for its impact on the long term value of $\pone$.  But,  the available data in India does not allow for a quantitative analysis of any of these factors, and we are not aware of any such analysis having been performed. So, at the moment, the comparison of the positive and negative effects of the lockdown must necessarily rely on a qualitative, and not a quantitative, analysis.

\section{Conclusion and discussion}

An analysis of simple epidemiological models, performed in section \ref{sectheory} and section \ref{secheuristic}, would suggest that the Indian lockdown has largely delayed deaths, and averted only a small number of deaths.

Nevertheless, some groups have attempted to claim that the lockdown has averted a significant number of deaths by using SEIR-models or more-general compartmental models. These estimates appear to compare fatalities between scenarios by an arbitrarily chosen fixed date.  

We showed that such claims are misleading and appear to arise from a lack of appreciation of some elementary aspects of the dynamics of such models. We emphasize that while our precise formulas were derived in the context of SEIR models in section \ref{sectheory}, our heuristic arguments from section \ref{secheuristic} are robust and also apply to more complicated models.

Simple epidemiological models do not account for several important real-world effects of the lockdown. This means that question of whether the Indian lockdown has really averted deaths must rely on a comparison of the effects described in section \ref{secrealworld}. A quantitative analysis of these effects is not possible given the extant data. 

So the factors listed in section \ref{secrealworld} must be compared qualitatively. In this paper, we do not provide a detailed qualitative analysis of these factors.
Nevertheless, to this author, it appears clear that the Indian lockdown has not led to any dramatic improvement in the ability of the healthcare system to address the COVID-19 pandemic. 
And the steps that have been taken to ramp up the production of Personal Protective Equipment (PPE) \cite{koshyhindu2020}, or increase testing \cite{neetuchandramint2020} appear to be far  from sufficient to offset the humanitarian and healthcare crises caused by the lockdown, or even the negative impact that the lockdown has had on the ability of people to sustainably maintain physical-distancing measures in the long run. Therefore, this author is of the opinion that the lockdown of the country has exacerbated the challenges of the COVID-19 pandemic and created the risk of excess deaths, rather than averting deaths.

Viable alternatives to a nationwide lockdown, which were advocated by several public health experts \cite{koshyhindu2020}, may have included early testing and tracing, localized lockdowns, and a  promotion of more sustainable physical-distancing measures, together with a strong focus on the welfare of people affected by restrictions.

\ifthenelse{\boolean{elaboratedeclare}}{
\section*{List of abbreviations}
\begin{enumerate}
\item
COVID-19: Coronavirus disease.
\item
SEIR: Susceptible-Exposed-Infectious-Recovered.
\item
BCG: Boston Consulting Group.
\item
TB: Tuberculosis.
\item
SWAN: Stranded Workers Action Network.
\item
PPE: Personal Protective Equipment.
\end{enumerate}

\section*{Declaration}
\paragraph{Ethics approval and consent to participate.} Not applicable.
\paragraph{Consent for publication.} Not applicable.
\paragraph{Availability of data and material.} All data used in the manuscript is cited where used and publicly available.
\paragraph{Competing interests.} The author declares that he has no competing interests.
\paragraph{Funding.} Not applicable.
\paragraph{Authors' Contribution.} This is a single author manuscript with no collaborators.}{}
\paragraph{Acknowledgments.}
I am grateful to Pinaki Chaudhuri, Sourendu Gupta, Alok Laddha,  Gautam Menon, Subroto Mukerjee, Madhusudhan Raman, Joseph Samuel, Srikanth Sastry, Ramachandran Shankar, Rahul Siddharthan and Supurna Sinha for comments on a draft of this manuscript. I am also grateful to members of the Indian Scientists Response to COVID-19 collective and the Politically Mathematics collective for several discussions on related issues. 

\ifthenelse{\boolean{includeapp}}{
\appendix
\section*{Appendix}
\section{The INDSCI-SIM model \label{appisim}}
In this appendix, we discuss the  INDSCI-SIM model \cite{INDSCISIM-2020}, since this is an extended compartmental model and it was also used to generate one of the figures mentioned  in the Introduction. 

Although the predictions of  INDSCI-SIM have been discussed in the popular media, as of the time this paper was written, the modellers had released neither a detailed preprint describing the model and related calculations, nor the computer code for the model. In the discussion below, we are forced to rely on an online tool that the modellers created, and some summary slides of documentation that they have put up. Both of these are accessible through the \href{https://indscicov.in/for-scientists-healthcare-professionals/mathematical-modelling/indscisim/}{model's website}. We found this information insufficient to recreate, in detail, the various steps that led to the final figure for ``deaths averted'' due to the lockdown that the modellers announced. Therefore we have been forced to make plausible inferences in what follows.

\paragraph{\bf Some generalities \\}
The INDSCI-SIM model differs from the SEIR model, since it has 9 compartments instead of 4, and additionally envisions a division of the Indian population into state-level populations that interact with each other.
For instance, there are compartments corresponding to those who are ``asymptomatic'', ``pre-symptomatic'', ``mildly symptomatic'', ``hospitalized'' etc. 

While the dynamics of this model are more {\em complicated}, it would be erroneous to conclude that they are a more {\em accurate} representation of reality as we now explain.

Even for a single sub-population, the model appears to have at least 14 parameters. Some of these are clinical: a parameter for the fraction of asymptomatic individuals, another parameter for the fraction for those who are pre-symptomatic but have mild symptoms etc.  Some other parameters correspond to social dynamics:  a parameter controlling the rate of contacts between asymptomatic and susceptible individuals, another controlling the same rate for mildly symptomatic individuals etc.  If one consider multiple sub-populations, then there are additional parameters that control the contacts between these populations.

What is common to all these parameters is that they are poorly understood and have certainly not been reliably quantified. As is well known to modellers, given a surfeit of parameters, it is possible to fit a variety of data. Moreover, by tweaking these parameters, it is often easy to obtain a variety of predictions.

 Nevertheless, as we explained above, the result that ``a lockdown does not avert deaths, but only delays them,'' is rather robust. This result constitutes one of the {\em few robust predictions} made by models like INDSCI-SIM. This result is also mentioned in the documentation for the model (slide 19 in v1.0).

 We used the model's online tool on 4 June 2020 to verify this property of the model. We maintained the tool's default settings,  which seed the epidemic only in Maharashtra.  (We emphasize that these default parameters are {\em not realistic} and used only for purposes of illustration.)  

 We set the  epidemic to start on 1 March, and allowed it  to run for 300 days. We compared the following two scenarios. In the first scenario, no intervention was introduced. In the second scenario,  we set the intervention to what the tool calls a  ``simple lockdown'' beginning on day 20 and continuing till day 90.

We found that, without the lockdown, the number of deaths in Maharashtra by 25 December was 31,372. With the lockdown, the number of deaths by that date was 31,302. So, in this example, the INDSCI-SIM model itself ``predicts'' that a simple lockdown of 70 days will save 70 lives asymptotically.

On the other hand, if one were to do something arbitrary like comparing deaths by 15 May, then one would find in this example that, without the lockdown, there would be 8,400 deaths whereas the lockdown would bring this number down to 217. But, of course, it would be extremely misleading to conclude that the lockdown has ``averted more than 8,000 deaths''. Even in this model, it has only delayed these deaths and they rapidly catch up with the no-lockdown scenario.

\paragraph{\bf Perspective on lockdowns \\}
In spite of this, the modellers suggest that a lockdown may be effective in the long term. (Slides 19 and 20 in v1.0.)

This is because, apart from the parameters above, the model introduces additional parameters corresponding to an exponentially increasing ability of the state to test and quarantine infected individuals. 
Mathematically, this is done by inserting an exponentially decaying intensity of contacts between infected and susceptible individuals. If $\gamma_i$ are the parameters that control this intensity (which are generalizations of the parameter $\pone$ in our discussion of the SEIR model), the INDSCI-SIM model sets
\be
\gamma_i(t) = \gamma_i(0) e^{-{t \over \tau_i}}
\ee
with different time-constants, $\tau_i$,  for the different categories of infected individuals. 

Since the parameters in the model now have an {\em explicit} time-dependence, the analysis of section \ref{generalcompart} is no longer formally valid. In words, the assumption is that the period of the lockdown is used to {\em exponentially improve} the ability of the healthcare system to test and quarantine. And as we explained in section \ref{secrealworld}, it is true that, {\em in principle}, such an exponential improvement would lead to a long-term positive effect.  

There are two important issues. The first is that the model does not include parameters for any of the negative long-term impacts of the lockdown, but only this possible positive impact. So the benefits of a lockdown are almost inbuilt into this model unless one sets $\tau_i \rightarrow \infty$.

But a more serious problem is that  there is no good independent measure of how much the healthcare system (including testing) has improved during the lockdown. Therefore there is no good way to quantify the parameters $\tau_i$ that enter above.

\paragraph{\bf Prediction for deaths averted \\}

We now turn to the  INDSCI-SIM claim that the Indian lockdown averted between 8,000 to 32,000 deaths and explain why this conclusion is baseless. 

The recorded fatality data \cite{covidjhudb} for India is shown from 11 March (the data of the first fatality) to 15 May (the time frame for the analysis under consideration) in Figure \ref{figfatindia}.
\begin{figure}[!h]
\begin{center}
\includegraphics[height=0.3\textheight]{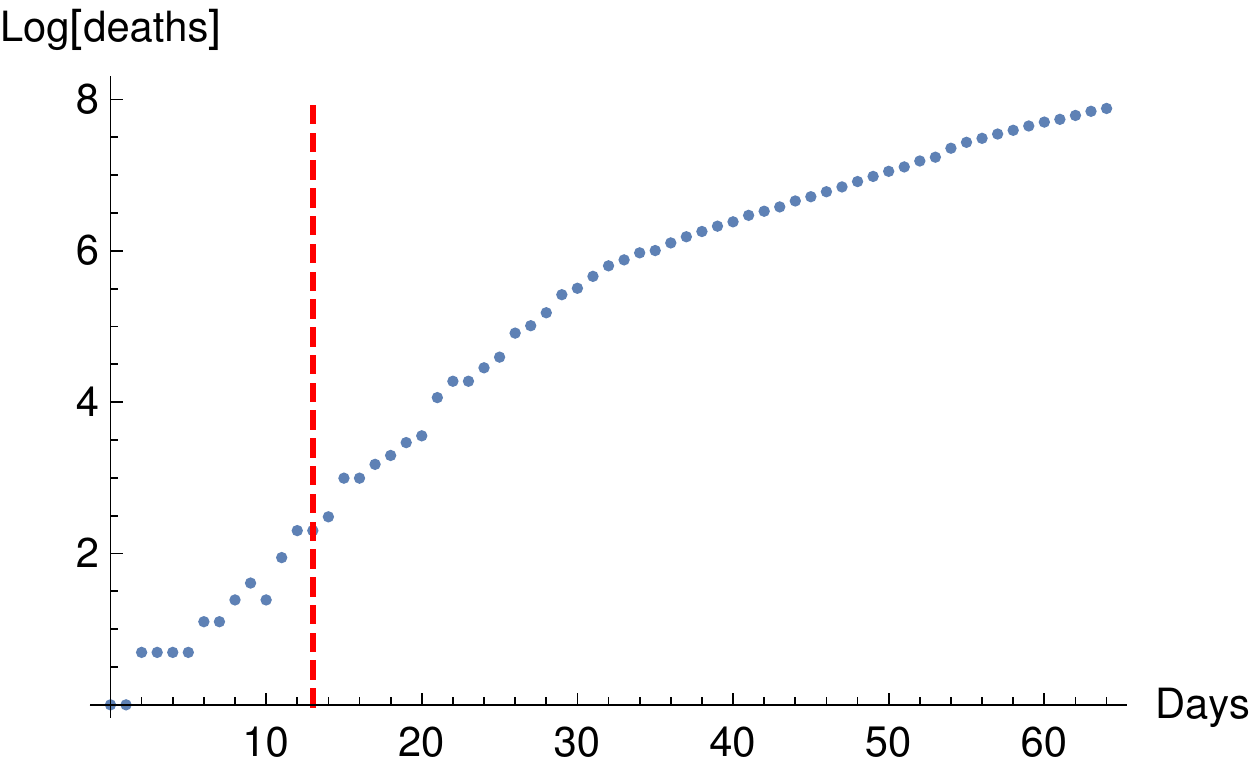}
\caption{\em Logarithm of the number of recorded deaths in India from 11 March to 15 May. The start of the lockdown is marked with a dashed line.\label{figfatindia}}
\end{center}
\end{figure}

The reader will immediately note that there are very few data points before the lockdown. In fact, the number of data points is smaller than the number of parameters for a single sub-population in the INDSCI-SIM model. 

Consequently, while the rate of growth of fatalities clearly slowed in April, there is {\em no way} to separate a number of effects above: the effect of the enforced physical distancing, the effect of increased testing-quarantining, the effect of behavioural changes and other precautions and the fact that the data might itself be flawed due to improvement in disease-surveillance over time \cite{gupta2020epidemic}. 
But this division is extremely important for understanding whether the lockdown has had a long-term impact since, as we have explained in great detail above: (a) the effects of enforced physical distancing due to the lockdown are transient (b) the effects of behavioural changes are long-lasting but not necessarily related to the lockdown (c) the effects of increased testing-quarantining are long-lasting.

Faced with this difficulty, the modellers of INDSCI-SIM appear to have adopted the following procedure, as far as we can discern from v1.2 of the documentation. They simply estimated a rate of exponential growth by fitting the data until April 3. (The significance of this particular date is unclear to us.) Then they extrapolated this rate of exponential growth to May 15, and compared it with the known cases on that date. They declared that the difference of these two figures was the number of ``deaths averted'' by the lockdown. Since the estimate of the  early exponential rate is prone to errors, which can have a significant impact over this time-period, the modellers obtained the large range of 8,000--32,000 ``deaths averted.''

We note two obvious points
\begin{enumerate}
\item
The idea that the calculation above yields the ``deaths averted'' {\em by} the lockdown is absurd. In fact, if the slowdown in the rate of growth in Figure \ref{figfatindia} arises from a combination of enforced physical distancing and behavioural changes (rather than increased testing-quarantining), then our analysis shows that the final number of deaths averted by the lockdown itself --- even before accounting for its other long-term negative impacts --- is likely to be very small.
\item
The simple analysis of extrapolating the rate of exponential growth beyond April 3 can be done without the INDSCI-SIM model, since any model would yield exponential growth in the early stages. Indeed, this procedure is probably the same as the procedure used by other groups mentioned  in the Introduction. Therefore, not only is the final figure misleading, it is also misleading to suggest that the computation has anything to do with a sophisticated epidemiological model. 
\end{enumerate}
}

\bibliography{refs}

\begin{thebibliography}{10}

\bibitem{Chandrashekhar_2020a}
V.~Chandrashekhar, ``1.3 billion people. a 21-day lockdown. can india curb the
  coronavirus?,'' {\em Science}, Mar 2020.
\newblock Available from: \url{http://dx.doi.org/10.1126/science.abc0030}.

\bibitem{Chandrashekhar_2020b}
V.~Chandrashekhar, ``As india’s lockdown ends, exodus from cities risks
  spreading covid-19 far and wide,'' {\em Science}, May 2020.
\newblock Available from: \url{http://dx.doi.org/10.1126/science.abc9596}.

\bibitem{kazminft2020}
{Amy Kazmin}, ``{Modi stumbles: India’s deepening coronavirus crisis },''
  {\em Financial Times}, 27 July 2020.
\newblock Available from:
  \url{https://www.ft.com/content/53d946cf-d4c2-4cc4-9411-1d5bb3566e83}
  [accessed 17 August 2020].

\bibitem{lockdown2020}
{ J. Jagannath}, ``{Lockdown saved 37,000-78,000 lives, averted 14 lakh-29 lakh
  Covid-19 cases: Govt},'' {\em Mint}, 22 May 2020.
\newblock Available from:
  \url{https://www.livemint.com/news/india/lockdown-saved-37-000-78-000-lives-averted-14-lakh-29-lakh-covid-19-cases-govt-11590147462140.html}
  [accessed 13 June 2020].

\bibitem{menon2020}
{Pinaki Chaudhuri} and {Gautam Menon}, ``{Explained: How many Covid-19 deaths
  prevented by lockdown?},'' {\em Indian Express}, 4 June 2020.
\newblock Available from:
  \url{https://indianexpress.com/article/explained/how-many-deaths-prevented-6441365/}
  [accessed 13 June 2020].

\bibitem{nairhindu2020}
{Sobhana K. Nair}, ``{Extended lockdown helped save nearly 78,000 lives, Health
  Ministry tells House committee },'' {\em The Hindu}, 4 August 2020.
\newblock Available from:
  \url{https://www.thehindu.com/news/national/extended-lockdown-helped-save-nearly-78-000-lives-health-ministry-tells-house-committee/article32271095.ece}
  [accessed 8 August 2020].

\bibitem{modijuneaddress2020}
{Narendra Modi}, ``{Address to the nation. Complete text},'' {\em Hindustan
  Times}, 30 June 2020.
\newblock Available from:
  \url{https://www.hindustantimes.com/india-news/pm-modi-s-address-to-the-nation-complete-text/story-pWz8lPwjANttEsMd47YDZI.html}
  [accessed 8 August 2020].

\bibitem{wu2013sensitivity}
J.~Wu, R.~Dhingra, M.~Gambhir, and J.~V. Remais, ``Sensitivity analysis of
  infectious disease models: methods, advances and their application,'' {\em
  Journal of The Royal Society Interface}, vol.~10, no.~86, p.~20121018, 2013.

\bibitem{bhatia2020lessons}
R.~Bhatia, P.~Abraham, {\em et~al.}, ``Lessons learnt during the first 100 days
  of covid-19 pandemic in india,'' {\em Indian Journal of Medical Research},
  vol.~151, no.~5, p.~387, 2020.

\bibitem{covidmediamay22}
{Press Information Bureau}.
\newblock ``{Media Briefing on COVID-19},'' [online].
\newblock 22 May 2020 [accessed 17 August 2020].
\newblock Available from:
  \url{https://twitter.com/PIB_India/status/1263804417020538881}.

\bibitem{allen2008mathematical}
L.~J. Allen, F.~Brauer, P.~Van~den Driessche, and J.~Wu, {\em Mathematical
  epidemiology}, vol.~1945.
\newblock Springer, 2008.

\bibitem{covidjhudb}
{CSEE at Johns Hopkins University}.
\newblock ``{CSSEGISandData},'' [online].
\newblock 2020 [accessed 16 June 2020].
\newblock Available from: \url{https://github.com/CSSEGISandData}.

\bibitem{economicsurvey1819}
{Ministry of Finance, Government of India}, ``Economic survey,'' 2019.
\newblock see vol. 2. Section on ``employment scenario''.
\newblock Available from:
  \url{https://www.thehinducentre.com/resources/article28283454.ece/binary/Economic\%20Survey\%20Volume\%20II\%20Complete\%20PDF.pdf}
  [accessed 17 August 2020].

\bibitem{cmiejun2020}
{Mahesh Vyas}.
\newblock ``{21 million jobs added in May},'' [online].
\newblock 2 June 2020 [accessed 17 June 2020].
\newblock Available from:
  \url{https://www.cmie.com/kommon/bin/sr.php?kall=warticle&dt=2020-06-02\%2011:43:41&msec=800}.

\bibitem{swan2020}
{Data Team}, ``{Data | 96\% migrant workers did not get rations from the
  government, 90\% did not receive wages during lockdown: survey},'' {\em The
  Hindu}, 20 April 2020.
\newblock Available from:
  \url{https://www.thehindu.com/data/data-96-migrant-workers-did-not-get-rations-from-the-government-90-did-not-receive-wages-during-lockdown-survey/article31384413.ece}
  [accessed 13 June 2020].

\bibitem{globaltbreport19}
{World Health Organization}, ``{Global Tuberculosis Report},'' 2019.
\newblock Available from:
  \url{https://apps.who.int/iris/bitstream/handle/10665/329368/9789241565714-eng.pdf}
  [accessed 17 August 2020].

\bibitem{beditb2020}
{Aneesha Bedi} and {Swagata Yadavar}, ``{TB patients badly hit by lockdown —
  80\% drop in diagnosis, huge struggle for medicines},'' {\em The Print}, 30
  April 2020.
\newblock Available from:
  \url{https://theprint.in/health/tb-patients-badly-hit-by-lockdown-80-drop-in-diagnosis-huge-struggle-for-medicines/411399/}
  [accessed 13 June 2020].

\bibitem{Baker_Wilson_Anglemyer_2020}
M.~G. Baker, N.~Wilson, and A.~Anglemyer, ``Successful elimination of covid-19
  transmission in new zealand,'' {\em New England Journal of Medicine},
  vol.~383, p.~e56, Aug 2020.
\newblock Available from: \url{http://dx.doi.org/10.1056/NEJMc2025203}.

\bibitem{Leung_Wu_Liu_Leung_2020}
K.~Leung, J.~T. Wu, D.~Liu, and G.~M. Leung, ``First-wave covid-19
  transmissibility and severity in china outside hubei after control measures,
  and second-wave scenario planning: a modelling impact assessment,'' {\em The
  Lancet}, vol.~395, p.~1382–1393, Apr 2020.
\newblock Available from:
  \url{http://dx.doi.org/10.1016/S0140-6736(20)30746-7}.

\bibitem{lee2020job}
K.~Lee, H.~Sahai, P.~Baylis, and M.~Greenstone, ``Job loss and behavioral
  change: The unprecedented effects of the india lockdown in delhi,'' {\em
  University of Chicago, Becker Friedman Institute for Economics Working
  Paper}, no.~2020-65, 2020.
\newblock Available from:
  \url{https://bfi.uchicago.edu/wp-content/uploads/BFI_WP_202065.pdf} [accessed
  17 August 2020].

\bibitem{bar2020encouraging}
N.~Bar-Zeev and W.~J. Moss, ``Encouraging results from phase 1/2 covid-19
  vaccine trials,'' {\em The Lancet}, vol.~396, pp.~448--449, 2020.
\newblock Available from:
  \url{https://www.thelancet.com/journals/lancet/article/PIIS0140-6736(20)31611-1/fulltext}.

\bibitem{koshyhindu2020}
{Jacob Koshy}, ``{Coronavirus | Enforcing lockdown indefinitely too disruptive,
  say public health experts },'' {\em The Hindu}, 1 June 2020.
\newblock Available from:
  \url{https://www.thehindu.com/news/national/coronavirus-indefinite-lockdown-more-harmful-say-public-health-experts/article31715670.ece}
  [accessed 9 August 2020].

\bibitem{neetuchandramint2020}
{Neetu Chandra Sharma}, ``{Covid-19 testing capacity to reach 10 lakh tests per
  day: PM Modi },'' {\em The Mint}, 27 July 2020.
\newblock Available from:
  \url{https://www.livemint.com/news/india/covid-19-testing-capacity-to-reach-10-lakh-tests-per-day-pm-modi-11595868800605.html}
  [accessed 17 August 2020].

\bibitem{INDSCISIM-2020}
{Snehal Shekatkar}, {Bhalchandra Pujari}, {Mihir Arjunwadkar}, {Dhiraj Kumar
  Hazra}, {Pinaki Chaudhuri}, {Sitabhra Sinha}, {Gautam I Menon}, {Anupama
  Sharma}, and {Vishwesha Guttal}, ``{INDSCI-SIM A state-level epidemiological
  model for India},'' 2020.
\newblock Available from: \url{https://indscicov.in/indscisim} [accessed 13
  June 2020].

\bibitem{gupta2020epidemic}
{Sourendu Gupta}, ``Epidemic parameters for covid-19 in several regions of
  india,'' May 2020.
\newblock arXiv:2005.08499.
\newblock Available from: \url{http://www.arxiv.org/abs/2005.08499}.

\end{thebibliography}
\bibliographystyle{ieeetrurl}

\end{document}